\documentclass[conference]{IEEEtran}
\IEEEoverridecommandlockouts

\usepackage{cite}
\usepackage{amsmath,amssymb,amsfonts}
\usepackage{algorithmic}
\usepackage{graphicx}
\usepackage{xcolor}
\usepackage{booktabs}
\usepackage{multirow}
\usepackage{array}
\usepackage{url}

\def\BibTeX{{\rm B\kern-.05em{\sc i\kern-.025em b}\kern-.08em
    T\kern-.1667em\lower.7ex\hbox{E}\kern-.125emX}}

\begin{document}

\title{MobiBench: Benchmarking LLMs for On-Device Performance}

\author{

\IEEEauthorblockN{Arya Hariharan}
\IEEEauthorblockA{\textit{Dept. of Computer Science}\\
\textit{and Engineering}\\
\textit{RV College of Engineering}\\
Bengaluru, India\\
aryahariharan.cs22@rvce.edu.in       }
\and

\IEEEauthorblockN{Rohit Suresh}
\IEEEauthorblockA{\textit{Department of Computer Science}\\
\textit{and Engineering}\\
\textit{RV College of Engineering}\\
Bengaluru, India\\
rohitsuresh.cs22@rvce.edu.in}
\and

\IEEEauthorblockN{Bolla Sai Naga Yashwanth}
\IEEEauthorblockA{\textit{Dept. of Information Science}\\
\textit{and Engineering}\\
\textit{RV College of Engineering}\\
Bengaluru, India\\
bollasainagay.is22@rvce.edu.in      }

\and

\IEEEauthorblockN{Ashok Senapati}
\IEEEauthorblockA{\textit{Samsung Research R\&D}\\
Bengaluru, India\\
a.ashokkumar@samsung.com}
\and
\IEEEauthorblockN{Thummala Pallavi}
\IEEEauthorblockA{\textit{Samsung Research R\&D}\\
Bengaluru, India\\
t.pallavi@samsung.com}

\and

\IEEEauthorblockN{Anala M R}
\IEEEauthorblockA{\textit{Dept. of Information Science}\\
\textit{and Engineering}\\
\textit{RV College of Engineering}\\
Bengaluru, India\\
analamr@rvce.edu.in}
\and
\IEEEauthorblockN{Soumya A}
\IEEEauthorblockA{\textit{Dept. of Computer Science}\\
\textit{and Engineering}\\
\textit{RV College of Engineering}\\
Bengaluru, India\\
soumyaa@rvce.edu.in}

}
\maketitle

\begin{abstract}
Understanding how large language models (LLMs) perform under real-world resource constraints has become more crucial due to the growing demand for deploying LLMs on mobile and edge devices. While there are a number of execution frameworks for on-device inference, the evaluation of llama.cpp as the runtime is the sole focus of this work. This work, presents a unified benchmarking suite that uses llama.cpp to thoroughly evaluate the performance of edge-optimized LLMs in a variety of hardware settings. The benchmark includes both user-facing metrics (task-specific accuracy, prefill speed, decode speed, time-to-first-token, etc.) and system-level measurements (memory consumption, battery consumption, etc.) and covers a fairly varied set of representative natural language tasks, such as summarization and question-answering. This paper present a comparative analysis that identifies important trade-offs resulting from model architecture and hardware features by methodically assessing several lightweight models on commercial system-on-chips (SoCs) and CPU/GPU platforms. In order to guide future developments in model optimization, runtime development, and edge-device hardware design for effective large-scale language intelligence, the benchmark lays the groundwork for repeatable and thorough evaluation of on-device LLMs using a single, widely used runtime.
\end{abstract}


\section{Introduction}
Through applications in question answering, summarization, dialogue systems, and code generation, among other areas~\cite{han2016deep, frantar2022gptq}, large language models (LLMs) have quickly developed into potent tools for both natural language understanding and generation. Due to their high computational requirements, these models have historically run on cloud servers. However, interest in on-device deployments of LLMs has increased due to growing concerns about latency, privacy, and personalization~\cite{lin2023awq}. Users have more control over their personal information when the inference is run locally, and applications become more responsive and less reliant on network connectivity.

Despite this, it remains difficult to deploy such LLMs on edge devices. Such system-on-chips (SoCs) are inherently resource-constrained, with strict bounds on memory, energy, and processing power~\cite{aiedgetorch2023}. In response, researchers and practitioners have increasingly turned to model compression methods such as quantization, pruning, and distillation, as well as lightweight architectures tailored for edge hardware~\cite{pytorch2023executorch, narayanan2021efficient}. Concurrently, optimized inference runtimes such as llama.cpp have been developed that are targeted at enabling efficient execution of LLMs across various hardware platforms. These efforts have made on-device LLMs more practical; however, single runtimes still have limited understanding of how model performance will vary with different tasks, hardware configurations, and optimization strategies.

Benchmarking is important to advance both system design and model development. While prior work has focused on benchmarking LLMs for either server-class environments~\cite{shoeybi2019megatron}, or for specific mobile workloads~\cite{mattson2020mlperf}, comprehensive benchmarks for on-device performance of LLMs on a diverse natural language tasks and SoCs do not exist. Second, without these benchmarks, it remains hard for developers to find bottlenecks, make effective trade-offs, or design new efficient applications that fully leverage on-device inferencing for LLMs.

To address this, the paper proposes a systematic benchmarking solution for evaluating LLMs for on-device computing scenarios using llama.cpp, an edge-specific runtime engine with capabilities to deploy on a range of hardware devices. Our proposed benchmark covers a broad range of natural language (NL) applications, including but not limited to summary and question-answer applications, and seeks to analyze both application-level performance metrics (such as latency, throughput, and accuracy) and system-level aspects (such as memory efficiency, quantization behavior, and energy efficiency). The paper aims to comparatively analyze open-source and lightweight LLM models on both CPU and GPU platforms, and aims to understand the important trade-offs between accuracy, efficiency, and resource usage for inferencing LLMs in constrained environments.

\section{Related Work}

The deployment of large language models (LLMs) on mobile and edge platforms has drawn significant interest in recent years, motivating research across model compression, efficient runtimes, and benchmarking. This section discusses the related work in which relevant studies are surveyed and identified to highlight gaps that are addressed in this research work.

Model compression has proven to be an essential enabler for the execution of on-device LLM inference. Traditional model compression techniques, such as Deep Compression~\cite{han2016deep}, used pruning and quantization to compress the model without affecting its performance. More recently, the need for model quantization has gotten considerable attention for the execution of on-device LLM inference, and GPTQ~\cite{frantar2022gptq} and AWQ~\cite{lin2023awq} provide the results to prove that 4-bit model quantization has the potential to save significant model storage and inference overhead without affecting quality. These methods are responsible for the execution of billion-parameter models on devices.

In addition to the compression techniques, light-weight execution frameworks have been developed to support the low-latency inference of LLMs with mobile CPUs, GPUs, and NPUs. \texttt{llama.cpp} has gained popularity due to its portability and support for quantized models in an optimized C++ runtime. AI EdgeTorch~\cite{lin2023awq} has been introduced as a PyTorch-integrated framework specifically designed for mobile platforms. It enables researchers to implement quantization-aware inference. Similarly, ExecuTorch~\cite{pytorch2023executorch} has been introduced for PyTorch to support quantized inference in an optimized runtime for various edge platforms.

Although the process of benchmarking has played an essential role in the evolution of machine learning algorithms in the cloud and server environments ~\cite{narayanan2021efficient,shoeybi2019megatron}, the state of LLM assessment in an on-device setting has remained in its infancy. Current benchmarks in the industry, including MLPerf ~\cite{mattson2020mlperf}, are more focused on training and inference in the datacenter paradigm. Recent studies such as EdgeFormer ~\cite{xu2023edgeformer} have started exploring its possibilities in mobile transformer networks. To date, there has been a lack of research dedicated to thoroughly benchmarking inference frameworks for on-device performance.

Research works have explored the influence of hardware-related aspects like dynamic voltage and frequency scaling (DVFS), memory bandwidth, and power constraints with regard to deep learning tasks~\cite{lane2016deepx,li2020edgeai}. These works highlight the need for hardware-software co-design, especially considering the fact that LLMs are more demanding compared to conventional on-device deep learning tasks like image classification. However, most of these works exist before the proliferation of LLMs; therefore, they do not focus on the computation patterns associated with LLMs.

To conclude, while research has prepared the ground for efficient on-device inference via compression approaches, specialized runtimes, and profiling at a systems level, there lacks an end-to-end benchmark that systematically compares and contrasts contemporary on-device LLM frameworks on various tasks and hardware substrates. This paper bridges this gap to provide a benchmark that brings much-needed reproducibility to comparisons of LLM deployment on resource-constrained devices and informs future model/runtime/hardware co-design studies.

\section{Experimental Setup}
Our experimental setup is designed to rigorously evaluate the performance of LLMs on-device. The setup aims to provide a unified framework for assessing runtime efficiency, accuracy, and resource trade-offs.

\subsection{Hardware Platforms}
To capture performance on various SoCs and CPU/GPU configurations, the evaluation is carried out on standard consumer hardware. The benchmarking setup includes:
\begin{itemize}
    \item A gaming laptop equipped with an Intel Core i7 12th-generation processor and an NVIDIA RTX 3060 GPU.
    \item A standard laptop featuring an Intel Core i5 processor with no dedicated graphics hardware.
\end{itemize}

\subsubsection{Benchmarking Frameworks}

All experiments in this work are performed exclusively using \texttt{llama.cpp}, a widely adopted lightweight inference runtime for deploying LLMs on CPUs, GPUs, and mobile-class devices. \texttt{llama.cpp} supports multiple hardware targets, including x86, ARM, Metal, BLAS, BLIS, SYCL, MUSA, CUDA, HIP, CANN, OpenCL, RPC, and Vulkan (version 1.2 or greater)~\cite{arm_llamacpp}. 

\begin{figure}[htbp]
    \centering
    \includegraphics[width=0.95\linewidth]{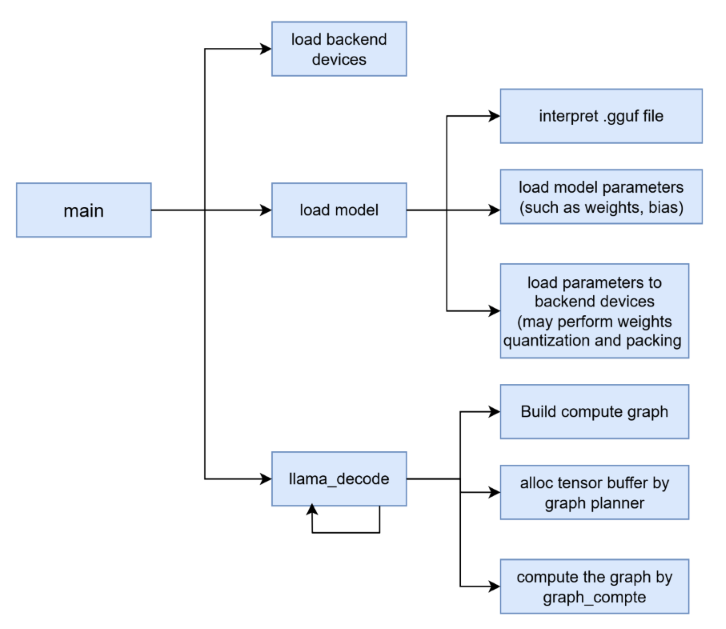}
    \caption{High-level execution pipeline of \texttt{llama.cpp}}
    \label{fig:llamacpp_structure}
\end{figure}

Figure~\ref{fig:llamacpp_structure} illustrates the high-level execution flow of \texttt{llama.cpp}, showing how the runtime initializes backend devices, loads and prepares model parameters from the \texttt{.gguf} format, constructs the compute graph, and executes iterative decoding during inference~\cite{arm_llamacpp}.

\subsubsection{Models Evaluated}\label{subsubsec2}
The benchmark includes several open-source, lightweight LLMs that are well-suited for on-device inferenceing due to their compact size and efficient architectures:

\begin{itemize}
    \item \textbf{Gemma (1B)}:  
    Gemma is a family of open-weight language models released by Google, designed to provide strong general-purpose language understanding while remaining computationally efficient. The 1B variant is optimized for low-latency inference and reduced memory usage, making it suitable for deployment on resource-constrained mobile and edge devices [17].

    \item \textbf{TinyLLaMA}:  
    TinyLLaMA is a compact transformer-based language model trained to replicate the behavior of larger LLaMA models at a significantly smaller scale. With approximately one billion parameters, it is commonly used for benchmarking and experimentation in constrained environments, offering a favorable trade-off between performance and efficiency [18].

    \item \textbf{LLaMA (3B)}:  
    LLaMA 3B is part of Meta’s LLaMA family of foundation language models designed for efficient training and inference. The 3B parameter variant provides improved semantic understanding and generation quality compared to smaller models, while remaining feasible to deploy on consumer-grade GPUs and edge platforms [19].

    \item \textbf{Phi-2}:  
    Phi-2 is a compact language model developed by Microsoft, trained using high-quality and carefully curated datasets with a focus on reasoning and instruction-following capabilities. Despite its relatively small size, Phi-2 demonstrates strong performance on reasoning and summarization tasks, making it well-suited for efficient on-device evaluation [20].
\end{itemize}

\subsubsection{Tasks and Datasets}
To evaluate model performance on practical tasks, we use curated subsets of four commonlu used benchmarks, capturing reasoning, factual recall, and summarisation.  

RepLiQA is an evaluation dataset that contains Context-Question-Answer triplets, where contexts are non-factual but natural-looking documents about made up entities such as people or places that do not exist in reality [14]. The CNN DailyMail Dataset is an English-language dataset containing just over 300k unique news articles as written by journalists at CNN and the Daily Mail [12] and is commonly used for summarization. The SciQ dataset contains 13,679 crowdsourced science exam questions about Physics, Chemistry and Biology, among others. The questions are in multiple-choice format with 4 answer options each [15]. Massive Multitask Language Understanding (MMLU) is a massive multitask test consisting of multiple-choice questions from various branches of knowledge. The test spans subjects in the humanities, social sciences, hard sciences, and other areas [13]. To create the test suite for the benchmark, 1000 questions were randomly sampled from each of these datasets. For the SciQ dataset, the official 1,000-entry test set was used instead of sampling. The dataset also recorded the ground-truth answer for each question.

\subsubsection{Evaluation Metrics}
We evaluate models along two primary axes: application-level performance and system-level efficiency.

\textbf{Application-Level Metrics:}
\begin{itemize}
    \item Speed: Prefill and decode speed (tokens/sec).
    \item Latency: Time-to-first-token (sec).
    \item Accuracy: Task-specific scores (e.g., ROUGE for summarization, accuracy for QA).
\end{itemize}

\textbf{System-Level Metrics:}
\begin{itemize}
    \item Memory: Peak RAM usage (measured in MB). This is collected using psutil to monitor the process's Resident Set Size (RSS).
    \item GPU: Peak GPU memory and average GPU utilization, collected via pynvml when inferencing on GPU.
    \item Efficiency: Quantization efficiency and energy consumption (Joules).
\end{itemize}
Task-specific performance metrics were also measured under for every combination of model, device, and task. For reproducibility, performance is averaged over several runs.

\section{Benchmark Evaluation}

\begin{figure}[htbp]
    \centering
    \includegraphics[width=\linewidth]{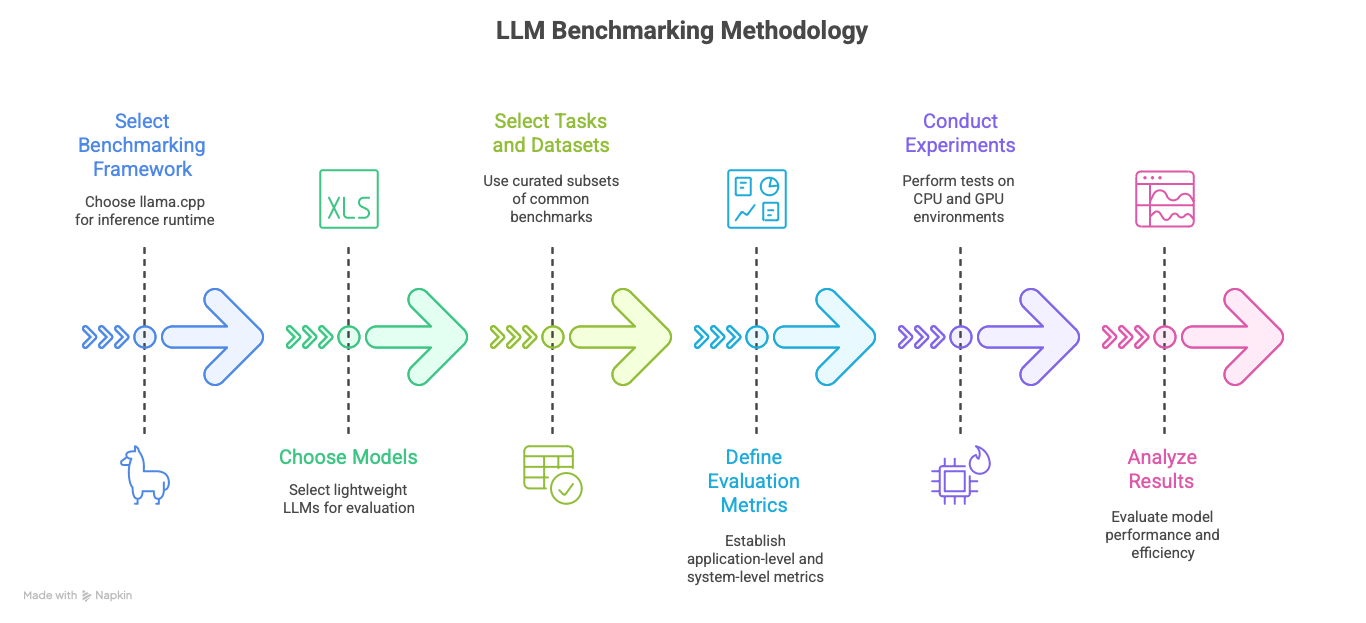}
    \caption{Overview of the LLM benchmarking methodology adopted in this work}
    \label{fig:benchmarking_methodology}
\end{figure}

Benchmark evaluations were performed by running the dataset on the designated hardware setups, with all outputs and timing measurements logged for analysis, following which the evaluation was conducted. Figure~\ref{fig:benchmarking_methodology} provides a high-level view of the benchmarking pipeline followed in this study, from selecting the inference framework and models to executing experiments and analyzing performance results.

For the summarization task, evaluations are done using contextual embeddings from a pretrained BERT model which were used to compare each model-generated summary with its matching gold reference in order to calculate BERTScore. The bert\_score function, which encodes both sets of sentences and calculates token-level similarity using cosine similarity in the embedding space, receives all gold and predicted texts in this evaluation. Recall quantifies how much of the reference meaning is captured by the prediction, precision indicates how well the predicted tokens are semantically aligned with the reference, and the F1 score gives their harmonic average. The final BERTScore metrics, which provide a reliable semantic-quality measure for the summarization outputs, are the mean precision, recall, and F1 across all assessed samples.

\subsection{Performance of Gemma 1B Model}
The effects of hardware acceleration on inference performance are shown in Table~\ref{tab:gemma1b_perf}. With wallclock durations ranging from 5.83 s (SciQ) to 27.95 s (RepLiQA), inference times on the CPU are noticeably longer. The expected throughput for CPU-based processing is in line with the prefill TPS and decode TPS values ($approx$ 50–60 and 20–23 tokens/s, respectively). CPU utilization is still high ($\sim$190--205\%), suggesting effective use of several threads. Optimized quantized model handling is suggested by the consistent memory usage of about 1 GB.

On the other hand, all datasets show significant speed gains in the GPU environment. While prefill throughput increases exponentially, surpassing 4000 tokens/s for RepLiQA and 3100 tokens/s for Summarization, average wallclock times decrease to roughly 2.2–2.5 seconds. Additionally, compared to CPU runs, decode throughput increases by almost 7–8$\times$. GPU runs show offloading of computation to CUDA cores while maintaining a moderate CPU usage ($\sim$99\%). GPU caching and model allocation overheads cause a slight increase in memory consumption ($sim$1.3 GB).

These results show that GPU acceleration considerably reduces latency and boosts token throughput without compromising memory efficiency or model stability. The consistency in throughput across all datasets shows that the Gemma 1B model maintains predictable performance under a variety of workloads and scales well on heterogeneous hardware environments.

\begin{table*}[htbp]
\centering
\caption{Performance Comparison of Gemma 1B Model on CPU and GPU Processors}
\label{tab:gemma1b_perf}
\scriptsize
\begin{tabular}{llcccccc}
\toprule
\textbf{Processors} & \textbf{Dataset} & \textbf{Samples} & \textbf{Time (s)} & \textbf{Prefill TPS} & \textbf{Decode TPS} & \textbf{CPU (\%)} & \textbf{Mem (MB)} \\
\midrule
\multirow{4}{*}{CPU} 
& RepLiQA & 638 & 27.95 & 52.53 & 19.97 & 197.16 & 1042.65 \\
& MMLU & 996 & 9.21 & 56.90 & 19.45 & 192.48 & 1009.04 \\
& SciQ & 1000 & 5.83 & 62.02 & 22.66 & 186.74 & 997.16 \\
& Summarization & 1367 & 22.61 & 57.60 & 20.57 & 205.60 & 1040.71 \\
\midrule
\multirow{4}{*}{GPU} 
& RepLiQA & 1000 & 2.21 & 4407.22 & 150.22 & 99.06 & 1309.43 \\
& MMLU & 1000 & 2.43 & 785.69 & 166.88 & 98.99 & 1311.51 \\
& SciQ & 1000 & 2.47 & 433.29 & 166.31 & 99.23 & 1311.28 \\
& Summarization & 1002 & 2.52 & 3102.01 & 161.15 & 98.83 & 1313.09 \\
\bottomrule
\end{tabular}
\end{table*}

The performance of the Gemma 1B model on three tasks Summarisation, MMLU, and SciQ on GPU and CPU using llama.cpp is summarized in Table~\ref{tab:gemma1b_acc}. With an average BERTScore F1 of 0.8417 (GPU) and 0.8457 (CPU) for summarization, the model consistently produced results that showed similar summary quality in both environments. The accuracy on MMLU remained low (0.2290 GPU, 0.2339 CPU), indicating a restricted capacity for reasoning but consistent performance. The lower SciQ scores (0.0690 GPU, 0.0470 CPU) indicate a challenge with domain-specific factual reasoning. Overall, the results demonstrate Gemma 1B's dependability on summarization tasks, but its reasoning accuracy is still restricted. The efficiency of llama.cpp for lightweight inference is demonstrated by the small performance difference between CPU and GPU.

\begin{table*}[htbp]
\caption{Accuracy Metrics of Gemma 1B Model on CPU and GPU Processors}
\label{tab:gemma1b_acc}
\centering
\scriptsize
\begin{tabular}{llcccc}
\toprule
\textbf{Processors} & \textbf{Dataset} &
\textbf{Samples} &
\textbf{BERTScore Precision} &
\textbf{BERTScore Recall} &
\textbf{BERTScore F1} \\
\midrule

GPU & Summarization & 1002 & 0.8069 & 0.8799 & 0.8417 \\
CPU  & Summarization & 1367 & 0.8072 & 0.8883 & 0.8457 \\
\midrule

\textbf{Processors} & \textbf{Dataset} &
\textbf{Samples} &
\textbf{Skipped} &
\textbf{Matched} &
\textbf{Accuracy} \\
\midrule

GPU & MMLU & 1000 & 12 & 229 & 0.2290 \\
CPU  & MMLU & 996 & 0 & 233 & 0.2339 \\
\midrule

GPU & SciQ & 1000 & 0 & 69 & 0.0690 \\
CPU  & SciQ & 1000 & 0 & 47 & 0.0470 \\
\midrule

GPU & RepLiQA & 1000 & 0 & 143 & 0.143 \\
CPU  & RepLiQA & 1000 & 0 & 231 & 0.231 \\
\bottomrule
\end{tabular}
\end{table*}

\subsection{Performance of TinyLlama Model}
Across all assessed datasets, the results in Table~\ref{tab:tinyllama_perf} show a notable performance advantage on the GPU over the CPU. Wall-clock times drop significantly from 37.58 s to 1.69 s on RepLiQA, 65.81 s to 1.39 s on MMLU, 18.70 s to 1.50 s on SciQ, and 21.81 s to 1.63 s on Summarization, with speedups ranging from 12 $\times$ to 47$\times$. While the GPU runs display 0.0 for these fields due to instrumentation differences rather than missing compute, the CPU runs report measurable prefill and decode throughput (20–85 TPS). Therefore, the most trustworthy foundation for comparison is wall-clock latency. Because of its shorter sequences and lower compute intensity, SciQ exhibits the smallest relative gain, whereas MMLU, which benefits from GPU-accelerated batched forward passes, shows the greatest improvement.

The platforms also differ in terms of memory usage and CPU utilization. While the GPU runs display $\sim$98\% host CPU usage while the majority of computation is offloaded to the GPU, the CPU runs exhibit multi-core saturation (187–209\% aggregate usage). In accordance with model weights and activations stored in GPU memory, peak host memory is marginally lower on the GPU ($\approx$1100--1145 MB) than on the CPU ($\approx$1300--1346 MB). Overall, the findings show that while CPU metrics are still helpful for comprehending token-level behavior, GPU execution offers significant latency and efficiency advantages for TinyLlama.

\begin{table*}[htbp]
\centering
\caption{Performance Comparison of TinyLlama Model on CPU and GPU Processors}
\label{tab:tinyllama_perf}
\scriptsize
\begin{tabular}{llcccccc}
\toprule
\textbf{Processors} & \textbf{Dataset} & \textbf{Samples} & \textbf{Time (s)} & \textbf{Prefill TPS} & \textbf{Decode TPS} & \textbf{CPU (\%)} & \textbf{Mem (MB)} \\
\midrule
\multirow{4}{*}{CPU} 
& RepLiQA & 1000 & 37.58 & 54.14 & 20.42 & 209.54 & 1346.58 \\
& MMLU & 996 & 65.81 & 75.04 & 27.90 & 187.29 & 1309.67 \\
& SciQ & 1000 & 18.70 & 85.41 & 29.70 & 190.50 & 1301.32 \\
& Summarization & 1002 & 21.81 & 65.98 & 24.27 & 205.01 & 1344.05 \\
\midrule
\multirow{4}{*}{GPU} 
& RepLiQA & 1000 & 1.69 & 0.0 & 0.0 & 97.63 & 1146.46 \\
& MMLU & 1000 & 1.40 & 0.0 & 0.0 & 98.44 & 1099.66 \\
& SciQ & 1000 & 1.50 & 0.0 & 0.0 & 98.60 & 1116.06 \\
& Summarization & 1002 & 1.63 & 0.0 & 0.0 & 98.23 & 1145.41 \\
\bottomrule
\end{tabular}
\end{table*}

Regardless of the underlying hardware, the accuracy evaluation of the TinyLlama model across CPU and GPU platforms, as displayed in Table~\ref{tab:tinyllama_acc}, shows that the model maintains highly consistent summarization quality and moderate reliability on multiple-choice reasoning tasks. Both platforms achieve comparable BERTScore metrics for the Summarization dataset, with GPU precision/recall/F1 of 0.8070/0.8799/0.8417 and CPU values of 0.8075/0.8880/0.8458. This suggests that generative quality is essentially constant across CPU and GPU inference. On the contrary aspect, task accuracy differences are more noticeable in structured evaluation settings. The CPU's accuracy of 0.2339 on MMLU is slightly higher than the GPU's accuracy of 0.191; this discrepancy can be partly explained by the GPU's 12 skipped or unparsable responses as opposed to the CPU's zero. Similar trends are seen in SciQ, where GPU accuracy is 0.018 and CPU accuracy is 0.011. However, rather than hardware effects, both scores represent the model's intrinsic limitations on scientific reasoning tasks. Overall, the benchmarking results demonstrate that TinyLlama's accuracy for generative tasks is generally consistent across hardware, whereas small variations in MCQ accuracy are more likely to result from variations in output formatting than from actual changes in model capability.

\begin{table*}[t]
\caption{Accuracy Metrics of TinyLlama Model on CPU and GPU Processors}
\label{tab:tinyllama_acc}
\centering
\scriptsize
\begin{tabular}{l l c c c c}
\toprule
\textbf{Processors} & \textbf{Dataset} &
\textbf{Samples} &
\textbf{BERTScore Precision} &
\textbf{BERTScore Recall} &
\textbf{BERTScore F1} \\
\midrule

GPU & Summarization & 1002 & 0.8070 & 0.8799 & 0.8417 \\
CPU  & Summarization & 1002 & 0.8075 & 0.8880 & 0.8458 \\
\midrule

\textbf{Processors} & \textbf{Dataset} &
\textbf{Samples} &
\textbf{Skipped} &
\textbf{Matched} &
\textbf{Accuracy} \\
\midrule

GPU & MMLU & 1000 & 12 & 191 & 0.1910 \\
CPU  & MMLU & 996 & 0 & 233 & 0.2339 \\
\bottomrule
\end{tabular}
\end{table*}

\subsection{Performance of Llama 3B Model}
The performance evaluation of the Llama 3B model on a GPU is shown in Table~\ref{tab:llama3b_perf}, which consistently demonstrates high efficiency across all four datasets, with average wall-clock times remaining within a narrow range of 1.96–2.17 seconds for 1000 input samples. With exceptionally high prefill throughput ranging from 450.57 TPS on SciQ to 4092.48 TPS on RepLiQA and robust decode throughput between 225 and 243 TPS, the GPU exhibits strong token processing capability, indicating rapid generation performance during both the prefill and autoregressive stages. The host CPU is actively coordinating inference while the GPU handles the majority of the computation, as evidenced by the CPU usage remaining steady at roughly 97–98\%. 
As would be expected for a 3B-parameter model loaded into GPU memory with little variation across tasks, peak memory consumption remains constant, ranging from 1339 MB to 1360 MB. Overall, these findings demonstrate that Llama 3B achieves high throughput, low latency, and stable resource utilisation on GPU hardware, which makes it ideal for real-time tasks like summarisation, multiple-choice reasoning, and question answering.

\begin{table*}[htbp]
\centering
\caption{Performance Comparison of Llama 3B Model on GPU Processor}
\label{tab:llama3b_perf}
\scriptsize
\begin{tabular}{lccccccc}
\toprule
\textbf{Dataset} & \textbf{Samples} & \textbf{Time (s)} & \textbf{Prefill TPS} & \textbf{Decode TPS} & \textbf{CPU (\%)} & \textbf{Mem (MB)} \\
\midrule
RepLiQA & 1000 & 2.18 & 4092.49 & 225.25 & 98.14 & 1360.11 \\
MMLU & 1000 & 1.96 & 830.24 & 237.85 & 97.98 & 1339.36 \\
SciQ & 1000 & 1.99 & 450.58 & 242.99 & 97.83 & 1341.96 \\
Summarization & 1002 & 2.17 & 3150.56 & 231.84 & 98.12 & 1360.30 \\
\bottomrule
\end{tabular}
\end{table*}

The accuracy evaluation of the Llama 3B model on GPU is shown in Table~\ref{tab:llama3b_acc}, which demonstrates strong summarisation quality and moderate performance on multiple-choice reasoning tasks. The model's high semantic alignment with reference outputs for summarisation is demonstrated by BERTScore values of 0.8002 precision, 0.8816 recall, and 0.8388 F1 across 997 evaluated samples of 1000 samples. These findings demonstrate that Llama 3B consistently preserves contextual coherence and meaning in the summaries it produces. However, performance on discrete evaluation tasks varies more: on SciQ, accuracy drops to 0.055 with two unparsable responses, while on MMLU, the model achieves an accuracy of 0.216 with only one output skipped. 
\begin{table*}[t]
\caption{Accuracy Metrics of LLaMA 3B Model on GPU Processor}
\label{tab:llama3b_acc}
\centering
\scriptsize
\begin{tabular}{l l c c c c}
\toprule
\textbf{Processor} & \textbf{Dataset} &
\textbf{Samples} &
\textbf{BERTScore Precision} &
\textbf{BERTScore Recall} &
\textbf{BERTScore F1} \\
\midrule

GPU & Summarization & 997 & 0.8002 & 0.8816 & 0.8388 \\
\midrule

\textbf{Processor} & \textbf{Dataset} &
\textbf{Samples} &
\textbf{Skipped} &
\textbf{Matched} &
\textbf{Accuracy} \\
\midrule

GPU & MMLU & 1000 & 1 & 216 & 0.2160 \\
GPU & SciQ & 1000 & 2 & 55 & 0.0550 \\
\bottomrule
\end{tabular}
\end{table*}

\subsection{Performance of Phi Model}
The performance of the Phi model on a GPU, with wall clock times ranging between 2.46 and 3.37 seconds, is reflected in Table ~\ref{tab:phi_perf}. The results signify good performance in terms of low latency for all four datasets. Strong prefill throughput is achieved by the GPU, which peaks at 2295.76 TPS on RepLiQA and stays above 478 TPS even for the most taxing tasks. Reliable token generation performance is demonstrated by the consistent maintenance of decode throughput between 102 and 122 TPS. CPU utilisation stays constant at 96–97\%, indicating that the GPU handles the majority of the computation while the host CPU actively manages inference. Peak memory usage, which ranges from 2070 to 2091 MB, is marginally higher than that of smaller models and is consistent with the memory footprint predicted for Phi's parameter size. 

\begin{table*}[htbp]
\centering
\caption{Performance Comparison of Phi Model on GPU Processor}
\label{tab:phi_perf}
\scriptsize
\begin{tabular}{lccccccc}
\toprule
\textbf{Dataset} & \textbf{Samples} & \textbf{Time (s)} & \textbf{Prefill TPS} & \textbf{Decode TPS} & \textbf{CPU (\%)} & \textbf{Mem (MB)} \\
\midrule
RepLiQA & 1000 & 2.73 & 2295.76 & 108.00 & 96.96 & 2090.39 \\
MMLU & 1000 & 3.37 & 698.59 & 102.30 & 97.08 & 2076.24 \\
SciQ & 1000 & 2.47 & 478.46 & 122.13 & 97.13 & 2070.85 \\
Summarization & 1002 & 3.15 & 1939.41 & 116.81 & 97.43 & 2091.31 \\
\bottomrule
\end{tabular}
\end{table*}

Table~\ref{tab:phi_acc} displays the Phi Model's accuracy evaluation on GPU, which shows good summarisation performance and moderate reliability on multiple-choice reasoning tasks. High BERTScore values of 0.8015 precision, 0.8823 recall, and 0.8398 F1 are obtained by the model for summarisation across 1000 of 1002 samples, demonstrating that the generated summaries maintain important semantic content and stay closely aligned with the reference texts. On the other hand, because of their discrete nature and more stringent correctness requirements, the structured evaluation tasks show lower accuracy. With 243 correctly matched outputs out of 1000 samples on MMLU, the model achieves an accuracy of 0.243. However, 16 outputs were skipped or unparsable, indicating sporadic formatting errors. For the more specialized SciQ dataset, performance drops to 0.082, with only 82 correct predictions and 13 skipped outputs, highlighting the model's challenges with domain-specific scientific reasoning.

\begin{table*}[t]
\caption{Accuracy Metrics of Phi Model on GPU Processor}
\label{tab:phi_acc}
\centering
\scriptsize
\begin{tabular}{l l c c c c}
\toprule
\textbf{Processor} & \textbf{Dataset} &
\textbf{Samples} &
\textbf{BERTScore Precision} &
\textbf{BERTScore Recall} &
\textbf{BERTScore F1} \\
\midrule

GPU & Summarization & 1000 & 0.8015 & 0.8823 & 0.8398 \\
\midrule

\textbf{Processor} & \textbf{Dataset} &
\textbf{Samples} &
\textbf{Skipped} &
\textbf{Matched} &
\textbf{Accuracy} \\
\midrule

GPU & MMLU & 1000 & 16 & 243 & 0.2430 \\
GPU & SciQ & 1000 & 13 & 82 & 0.0820 \\
\bottomrule
\end{tabular}
\end{table*}

\begin{table*}[t]
\caption{Performance Metrics of Phi2\_Q2\_K Model on GPU Processor}
\label{tab:phi2_perf}
\centering
\scriptsize
\begin{tabular}{l l c c c c c c}
\toprule
\textbf{Processor} & \textbf{Dataset} &
\textbf{Total Samples} &
\textbf{Avg Wallclock Time (s)} &
\textbf{Avg Prefill TPS} &
\textbf{Avg Decode TPS} &
\textbf{Avg CPU Usage (\%)} &
\textbf{Avg Peak Memory (MB)} \\
\midrule
GPU & Context\_QA     & 1000 & 3.8863 & 1465.2805 & 47.3969 & 96.7042 & 1702.3544 \\
GPU & MMLU\_mcq       & 1000 & 3.7512 & 691.6386  & 55.0112 & 97.0428 & 1699.1854 \\
GPU & Scientific\_mcq & 1000 & 3.6588 & 456.1964  & 56.2853 & 96.6729 & 1694.5500 \\
GPU & Summarization   & 1002 & 5.5051 & 1237.9824 & 46.6215 & 96.1103 & 1710.3262 \\
\bottomrule
\end{tabular}
\end{table*}

Quantized versions of the model were also evaluated, the results shown in Table~\ref{tab:phi2_perf} and Table~\ref{}. The average wallclock time varies from 3.65 seconds for \textit{Scientific\_mcq} to 5.50 seconds for \textit{Summarization}; the latter takes slightly longer due to its greater generation length. Due to variations in input complexity, prefill throughput varies by task, peaking at 
1465~tokens/s (\textit{Context\_QA}) and falling to 456~tokens/s (\textit{Scientific\_mcq}). Stable token generation performance is indicated by the decode throughput, which remains 
consistent at 46--56~tokens/s. During GPU execution, CPU utilization remains between 96 and 97 percent, indicating balanced  host coordination. A consistent peak memory usage of approximately 1.7~GB demonstrates 
effective quantized model handling. Overall, the \texttt{Phi2\_Q2\_K} model confirms dependable GPU-accelerated inference  performance across a range of workloads by delivering consistent latency, stable decode  speed, and predictable memory usage.

\begin{table*}[t]
\caption{Accuracy Metrics of Phi2\_Q2\_K Model on GPU Processor}
\label{tab:phi2_acc}
\centering
\scriptsize
\begin{tabular}{l l c c c c}
\toprule
\textbf{Operating System} & \textbf{Dataset} &
\textbf{Total Samples} &
\textbf{Avg BERTScore Precision} &
\textbf{Avg BERTScore Recall} &
\textbf{Avg BERTScore F1} \\
\midrule
GPU & Summarization & 1002 & 0.8013 & 0.8823 & 0.8398 \\
\midrule
\textbf{Operating System} & \textbf{Dataset} &
\textbf{Total Samples} &
\textbf{Skipped/Unparsed} &
\textbf{Matched Samples} &
\textbf{Accuracy} \\
\midrule
GPU & MMLU\_mcq       & 1000 & 0 & 233 & 0.2330 \\
GPU & Scientific\_mcq & 1000 & 0 & 69  & 0.1340 \\
\bottomrule
\end{tabular}
\end{table*}

The \texttt{Phi2\_Q2\_K} model's performance in summarization and multiple-choice tasks 
under GPU execution is shown by the accuracy results in Table~\ref{tab:phi2_acc}. For \textit{Summarization} (1002 samples), the model obtains a BERTScore Precision of 
0.8013, Recall of 0.8823, and F1 score of 0.8398. The higher recall relative to precision suggests that the generated summaries capture the majority of reference content, albeit with 
some redundancy or wording variation. The F1 score (${\approx}0.84$) indicates strong overall 
semantic alignment with ground-truth summaries. Performance is comparatively lower for tasks involving objective reasoning. On 
\textit{MMLU\_mcq} (1000 samples), the model achieves an accuracy of 23.3\% (233 correct), 
while accuracy falls to 13.4\% (69 correct) on \textit{Scientific\_mcq} (1000 samples). The absence of skipped or unparsable outputs indicates stable response formatting throughout 
evaluation. Overall, the \texttt{Phi2\_Q2\_K} model demonstrates solid semantic generation capability 
in summarization, but relatively limited factual and domain-specific reasoning accuracy 
across multiple-choice benchmarks.

The \texttt{Phi2\_Q3\_KM} model's inference performance under GPU acceleration across 
several datasets is shown in Table~\ref{tab:phi2_q3km_perf}. The average wallclock time ranges from 2.63 seconds (\textit{Scientific\_mcq}) to 
4.01 seconds (\textit{Summarization}), indicating generally faster execution compared 
to lower quantized variants. The longer generation sequences of summarization tasks 
account for the slightly increased wallclock time. Due to dataset-specific input 
characteristics, \textit{Context\_QA} achieves the highest prefill throughput at 
2171~tokens/s, followed by \textit{Summarization} at 1523~tokens/s, while 
\textit{Scientific\_mcq} records the lowest prefill rate at 452~tokens/s. Decode throughput shows clear improvement, reaching 87.99~tokens/s for 
\textit{Context\_QA} and 86.86~tokens/s for \textit{Scientific\_mcq}. However, due 
to longer output lengths, \textit{Summarization} yields a comparatively lower decode 
speed of 58.43~tokens/s. CPU utilization remains stable at approximately 96--97\%, 
demonstrating effective coordination between CPU and GPU resources. Peak memory usage 
increases to approximately 1.98--2.00~GB, reflecting the higher memory footprint 
associated with the Q3\_KM quantization level. Overall, the \texttt{Phi2\_Q3\_KM} model achieves reduced latency and improved decode 
throughput while maintaining stable resource utilization, demonstrating efficient 
GPU-accelerated performance across diverse NLP workloads.

\begin{table*}[t]
\caption{Performance Metrics of Phi2\_Q3\_KM Model on GPU Processor}
\label{tab:phi2_q3km_perf}
\centering
\scriptsize
\begin{tabular}{l l c c c c c c}
\toprule
\textbf{Processor} & \textbf{Dataset} &
\textbf{Total Samples} &
\textbf{Avg Wallclock Time (s)} &
\textbf{Avg Prefill TPS} &
\textbf{Avg Decode TPS} &
\textbf{Avg CPU Usage (\%)} &
\textbf{Avg Peak Memory (MB)} \\
\midrule
GPU & Context\_QA     & 1000 & 2.7711 & 2171.0830 & 87.9943 & 96.7152 & 2000.0834 \\
GPU & MMLU\_mcq       & 1000 & 2.8236 & 725.9316  & 75.5094 & 96.4092 & 1987.1585 \\
GPU & Scientific\_mcq & 1000 & 2.6272 & 451.5255  & 86.8588 & 96.7688 & 1982.4192 \\
GPU & Summarization   & 1002 & 4.0111 & 1522.8521 & 58.4312 & 96.6606 & 2000.6335 \\
\bottomrule
\end{tabular}
\end{table*}

\begin{table*}[t]
\caption{Accuracy Metrics of Phi2\_Q3\_KM Model on GPU Processor}
\label{tab:phi2_q3km_acc}
\centering
\scriptsize
\begin{tabular}{l l c c c c}
\toprule
\textbf{Operating System} & \textbf{Dataset} &
\textbf{Total Samples} &
\textbf{Avg BERTScore Precision} &
\textbf{Avg BERTScore Recall} &
\textbf{Avg BERTScore F1} \\
\midrule
GPU & Summarization & 1002 & 0.8016 & 0.8823 & 0.8399 \\
\midrule
\textbf{Operating System} & \textbf{Dataset} &
\textbf{Total Samples} &
\textbf{Skipped/Unparsed} &
\textbf{Matched Samples} &
\textbf{Accuracy} \\
\midrule
GPU & MMLU\_mcq       & 1000 & 0 & 263 & 0.2630 \\
GPU & Scientific\_mcq & 1000 & 0 & 255 & 0.2550 \\
\bottomrule
\end{tabular}
\end{table*}

The \texttt{Phi2\_Q3\_KM} model's performance on summarization and multiple-choice 
tasks under GPU execution is shown by the accuracy results in 
Table~\ref{tab:phi2_q3km_acc}. For \textit{Summarization} (1002 samples), the model obtains a BERTScore Precision 
of 0.8016, Recall of 0.8823, and F1 score of 0.8399. The higher recall relative to 
precision indicates strong content coverage, while the F1 score (${\approx}0.84$) 
reflects consistent semantic alignment with reference summaries. Overall, 
summarization quality remains stable and comparable to lower quantized variants. Performance improves notably for reasoning benchmarks. On \textit{MMLU\_mcq} 
(1000 samples), the model achieves an accuracy of 26.3\% (263 correct), outperforming 
Q2-level quantization. On \textit{Scientific\_mcq} (1000 samples), accuracy reaches 
25.5\% (255 correct), demonstrating a marked improvement in domain-specific reasoning 
ability. The absence of skipped or unparsable outputs confirms stable inference 
behaviour throughout evaluation. Overall, the \texttt{Phi2\_Q3\_KM} model demonstrates strong summarization performance 
alongside enhanced factual and multiple-choice reasoning accuracy, highlighting the 
advantages of the higher quantization configuration.

\section{LLM-Based Evaluation for Summarization Dataset}
\begin{figure}[htbp]
    \centering
    \includegraphics[width=\linewidth]{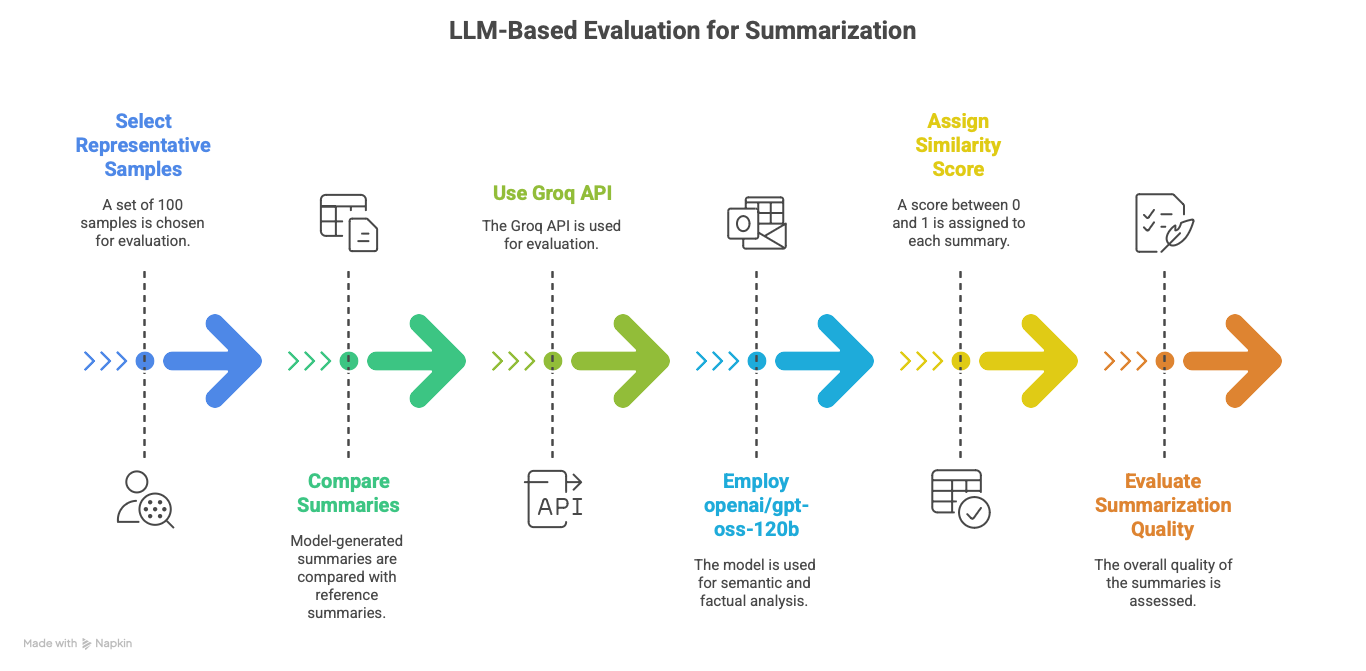}
    \caption{Workflow of the LLM-based evaluation framework for summarization}
    \label{fig:llm_eval_summarization}
\end{figure}

Using the Groq API, an evaluation framework based on the Large Language Model (LLM) was developed to evaluate the quality of the generated summaries. The model-generated summary and its matching gold (reference) summary were methodically compared for each of the 100 representative samples that were chosen. This evaluation performed semantic and factual analysis using the openai/gpt-oss-120b model hosted on the Groq platform in place of conventional lexical overlap metrics. In order to determine the degree of semantic alignment, meaning preservation, and factual consistency between the generated and reference summaries, the model was asked to provide a continuous similarity score in the range [0,1]. This method effectively captures subtleties that traditional metrics frequently miss, offering a more human-aligned, context-aware evaluation of summarisation quality. Groq's low-latency inference further enabled fast and scalable evaluation across all samples.
Figure~\ref{fig:llm_eval_summarization} illustrates sample selection, comparison between generated and reference summaries, semantic analysis using a large language model, similarity score assignment, and final quality assessment.
\subsection{Gemma 1B Model}
The results of the LLM-based evaluation conducted on 100 summarisation samples under two distinct execution environments-GPU and CPU-are shown in able~\ref{tab:llm_gemma}. The GPU configuration obtained a maximum score of 0.96, a mode of 0.45, a median of 0.5650, and an average similarity score of 0.5852. With a mean of 0.6961, a median of 0.75, a mode of 0.6, and a maximum of 1.0, the CPU configuration, on the other hand, generated higher overall evaluation values. According to the LLM, these findings show that the CPU environment produced slightly more factually correct and semantically consistent summaries. Small variations in runtime precision, hardware-level numerical computation, or environmental dependencies influencing model inference consistency could be the cause of the observed variation in scores across platforms. Overall, the higher central tendency measures observed for the CPU evaluation demonstrate improved alignment between predicted and reference summaries, suggesting a slightly more stable or deterministic performance of the summarization pipeline in this configuration.

\begin{table}[htbp]
\centering
\caption{LLM-based Evaluation Metrics of Gemma 1B Model}
\label{tab:llm_gemma}
\begin{tabular}{lcccc}
\toprule
\textbf{Processors} & \textbf{Mean} & \textbf{Median} & \textbf{Mode} & \textbf{Max} \\
\midrule
GPU & 0.5852 & 0.5650 & 0.45 & 0.96 \\
CPU & 0.6961 & 0.7500 & 0.60 & 1.00 \\
\bottomrule
\end{tabular}
\end{table}

\subsection{TinyLlama Model}
The LLM-based assessment of TinyLlama summarisation performance is shown in Table~\ref{tab:llm_tiny}, which indicates that both CPU and GPU runs generate semantic quality that is essentially similar, with only slight differences across the scoring metrics. While the median and mode scores are the same (0.45) on both platforms, the CPU run achieves a slightly higher mean score (0.5042) than the GPU (0.4554), suggesting that the central tendency of the semantic quality distribution is essentially unchanged. Additionally, the CPU achieves a slightly higher maximum score (0.9600 vs. 0.9200), indicating that some summaries produced by the CPU have better semantic alignment with the gold references.  These minute differences can be ascribed neither to the hardware capacity nor to the autoregressive decoding process itself but to minute variations that may occur during the sampling or execution processes in the autoregressive model. On the whole, the experiment shows that the semantic summarization quality of the model is comparable on the CPU as well as the GPU, thus establishes the negligible effect of the hardware environment on the linguistic quality of the model's outputs.

\begin{table}[htbp]
\centering
\caption{LLM-based Evaluation Metrics of TinyLlama Model}
\label{tab:llm_tiny}
\begin{tabular}{lcccc}
\toprule
\textbf{Processor} & \textbf{Mean} & \textbf{Median} & \textbf{Mode} & \textbf{Max} \\
\midrule
GPU & 0.4554 & 0.4500 & 0.45 & 0.9200 \\
CPU & 0.5042 & 0.4500 & 0.45 & 0.9600 \\
\bottomrule
\end{tabular}
\end{table}

\subsection{Llama 3B Model}
Table~\ref{tab:llm_llama3b} illustrates the evaluation of the LLaMA 3B model running on GPU with an average semantic score of 0.5075, reflecting its overall superior summarization capability. The median score of 0.55 and mode score of 0.55 illustrate how over half the summaries have moderately-to-strongly semantic alignment with the ground-truth summaries, which is an additional testament to the model's semantic alignment capabilities with the ground truths. The model sustains an overall strong semantic alignment with the ground truths with respect to the maximum score of 0.96, which reinforces how the summaries have strong semantic alignment with the actual ground-truth summaries generated during ground-truth summaries generation processes. The overall evaluation emphasizes how the summarization capabilities of the model are robust with respect to the semantic alignment evaluation process using the LLM-based semantic alignment model framework.

\begin{table}[htbp]
\centering
\caption{LLM-based Evaluation Metrics of Llama 3B Model}
\label{tab:llm_llama3b}
\begin{tabular}{lcccc}
\toprule
\textbf{Processor} & \textbf{Mean} & \textbf{Median} & \textbf{Mode} & \textbf{Max} \\
\midrule
GPU & 0.5075 & 0.5500 & 0.55 & 0.96 \\
\bottomrule
\end{tabular}
\end{table}

\subsection{Phi Model}
Table~\ref{tab:llm_phi} shows the assessment results for the Phi model on Windows-GPU for summarization ability, achieving an overall semantic score of 0.6056, which is the highest among the models compared. The median and mode for this value are both 0.65, emphasizing the fact that most of these summaries have outstanding semantic relevance compared to the gold standards. The maximum value for this assessment is 0.96, thereby emphasizing the fact that many of these summaries are nearly identical in meaning to the gold standards. These assessment results emphasize Phi's superior summarization abilities compared to the other models, which are small in size, thereby emphasizing the success of its model architecture.

\begin{table}[htbp]
\centering
\caption{LLM-based Evaluation Metrics of Phi Model}
\label{tab:llm_phi}
\begin{tabular}{lcccc}
\toprule
\textbf{Processor} & \textbf{Mean} & \textbf{Median} & \textbf{Mode} & \textbf{Max} \\
\midrule
GPU & 0.6056 & 0.6500 & 0.65 & 0.96 \\
\bottomrule
\end{tabular}
\end{table}

\begin{table}[t]
\caption{LLM-Based Evaluation Metrics of Phi2\_Q2\_K Model on GPU Processor}
\label{tab:phi2_llm}
\centering
\scriptsize
\begin{tabular}{l l c c c c}
\toprule
\textbf{Operating System} & 
\textbf{Mean} &
\textbf{Median} &
\textbf{Mode} &
\textbf{Max} \\
\midrule
GPU &  0.5085 & 0.5500 & 0.55 & 0.9600 \\
\bottomrule
\end{tabular}
\end{table}

The summaries produced by the \texttt{Phi2\_Q2\_K} model are evaluated for semantic 
quality using the LLM-based evaluation results shown in Table~\ref{tab:phi2_llm}. For the 100 assessed summarization samples, the model achieves a mean score of 0.5085, 
indicating a moderate level of overall semantic alignment with the reference summaries. The median score of 0.55 and mode of 0.55 suggest that the majority of summaries 
consistently fall within the mid-quality range, reflecting stable but not highly 
optimized semantic performance. The maximum score of 0.96 demonstrates that in several cases, the generated summaries 
were nearly identical in meaning and factual consistency to the gold references. 
Nevertheless, the gap between the mean and maximum score indicates variability in 
output quality across samples. Overall, the results show that the \texttt{Phi2\_Q2\_K} model produces reasonably 
coherent and semantically aligned summaries, with consistent mid-range performance 
and occasional high-quality outputs.

\begin{table}[t]
\caption{LLM-Based Evaluation Metrics of Phi2\_Q3\_KM Model on GPU Processor}
\label{tab:phi2_q3km_llm}
\centering
\scriptsize
\begin{tabular}{l l c c c c}
\toprule
\textbf{Operating System} & 
\textbf{Mean} &
\textbf{Median} &
\textbf{Mode} &
\textbf{Max} \\
\midrule
GPU & 0.6269 & 0.6500 & 0.65 & 0.9500 \\
\bottomrule
\end{tabular}
\end{table}

The \texttt{Phi2\_Q3\_KM} model's semantic performance on summarization tasks is 
reflected in the LLM-based evaluation results shown in Table~\ref{tab:phi2_q3km_llm}. The model achieves a mean score of 0.6269 across the evaluated samples, indicating 
stronger overall semantic alignment compared to lower quantized variants. The median 
score of 0.65 and mode of 0.65 suggest that the majority of summaries consistently 
fall within a higher quality range, reflecting stable and dependable output generation. The maximum score of 0.95 confirms that a number of generated summaries are nearly 
identical in meaning and factual consistency to the reference summaries. Furthermore, 
the comparatively narrow gap between the mean and median scores suggests balanced 
performance with reduced variability across samples.

Overall, the \texttt{Phi2\_Q3\_KM} model demonstrates a discernible improvement in 
summary quality and semantic similarity, illustrating how effectively the higher 
quantization configuration preserves generation fidelity.

\section{Conclusion}
A consistent and repeatable benchmarking framework for evaluating edge-optimized large language models on various hardware platforms is presented in this work. In a thorough comparison of the TinyLlama, Llama 3B, and Phi models on CPU and GPU environments, we found that while semantic output quality is generally consistent across hardware, runtime efficiency varies considerably. The benefits of hardware acceleration in low-latency inference were confirmed by GPU executions, which generally produced significant speedups in wall-clock time, prefill throughput, and decode throughput. Conversely, accuracy and LLM-based semantic evaluation metrics show slight differences between CPU and GPU outputs, indicating that hardware primarily affects performance rather than linguistic quality.

Llama 3B produced robust and dependable summarisation performance, TinyLlama produced lightweight, consistent results appropriate for constrained environments, and Phi had the best semantic alignment among the benchmarked models under LLM-based scoring. When combined, these results highlight the need for a standardised, task-aware, hardware-inclusive benchmarking methodology when implementing an LLM on an edge device. The suggested benchmarking suite makes use of a single runtime, llama.cpp, as a first step towards such a methodical investigation of on-device LLM inference. It offers insights and direction for future research on model compression, runtime optimisation, and effective deployment of LLMs on commercial edge hardware.


\begin{thebibliography}{00}

\bibitem{han2016deep} S. Han, H. Mao, and W. J. Dally, ``Deep compression: Compressing deep neural networks with pruning, trained quantization and Huffman coding,'' in \emph{ICLR}, 2016.

\bibitem{frantar2022gptq} E. Frantar, S. Ashkboos, T. Hoefler, and D. Alistarh, ``GPTQ: Accurate post-training quantization for generative pretrained transformers,'' in \emph{NeurIPS}, 2022.

\bibitem{lin2023awq} J. Lin et al., ``AWQ: Activation-aware weight quantization for LLM compression and acceleration,'' arXiv:2306.00978, 2023.

\bibitem{aiedgetorch2023} AI EdgeTorch, ``AI EdgeTorch: Efficient PyTorch runtime for mobile and embedded devices,'' Meta AI, 2023.

\bibitem{pytorch2023executorch} PyTorch Team, ``ExecuTorch: An execution runtime for edge AI,'' 2023.

\bibitem{narayanan2021efficient} D. Narayanan et al., ``Efficient large-scale language model training on GPU clusters,'' in \emph{OSDI}, 2021.

\bibitem{shoeybi2019megatron} M. Shoeybi et al., ``Megatron-LM: Training multi-billion parameter language models using model parallelism,'' arXiv:1909.08053, 2019.

\bibitem{mattson2020mlperf} P. Mattson et al., ``MLPerf: An industry standard benchmark suite for machine learning performance,'' in \emph{ISCA}, 2020.

\bibitem{xu2023edgeformer} S. Xu et al., ``EdgeFormer: Efficient transformer deployment for mobile and edge devices,'' in \emph{ICML}, 2023.

\bibitem{lane2016deepx} N. D. Lane et al., ``DeepX: A software accelerator for low-power deep learning inference on mobile devices,'' in \emph{IPSN}, 2016.

\bibitem{li2020edgeai} X. Li et al., ``Edge AI: On-demand accelerating deep neural network inference via edge computing,'' \emph{IEEE Trans. Mobile Computing}, 2020.

\bibitem{arm_llamacpp}
ARM,
``Understanding the \texttt{llama.cpp} Execution Pipeline,''
ARM Learn Documentation, 2023.



\bibitem{hermann2015teaching} K. M. Hermann et al., ``Teaching machines to read and comprehend,'' in \emph{NeurIPS}, 2015.

\bibitem{hendrycks2021measuring} D. Hendrycks et al., ``Measuring massive multitask language understanding,'' in \emph{ICLR}, 2021.

\bibitem{lewis2021paq} P. Lewis et al., ``PAQ: 65 million probably-asked questions and what you can do with them,'' \emph{TACL}, 2021.

\bibitem{welbl2017crowdsourcing} J. Welbl, N. F. Liu, and M. Gardner, ``Crowdsourcing multiple choice science questions,'' in \emph{Proc. 3rd Workshop on Noisy User-generated Text}, 2017.

\bibitem{llama}  Gerganov, Georgi; Nguyen, Xuan Son; Slaren (August 13, 2024). "Introduction to ggml". Huggingface.

\bibitem{ref17}
Google DeepMind, ``Gemma: Open Models Based on Gemini Research,'' 2024.
[Online]. Available: https://ai.google.dev/gemma

\bibitem{ref18}
P. Zhang et al., ``TinyLLaMA: An Open-Source Small Language Model,'' 2023.
[Online]. Available: https://github.com/jzhang38/TinyLlama

\bibitem{ref19}
H. Touvron \emph{et al.}, ``LLaMA: Open and Efficient Foundation Language Models,''
\emph{arXiv preprint arXiv:2302.13971}, 2023.

\bibitem{ref20}
Y. Li \emph{et al.}, ``Phi-2: Small Language Models with Strong Reasoning Capabilities,''
\emph{arXiv preprint arXiv:2309.05463}, 2023.


\end{thebibliography}
\end{document}